\documentclass[trackchanges]{aastex701}
\usepackage{subcaption}

\def\eqref#1{Equation~\ref{#1}} 
\def\revise#1{\textcolor{red}{\bf #1}}

\usepackage{comment}
\usepackage{xspace}
\def\src{V4641~Sgr}

\def\chandra{Chandra}
\def\nustar{NuSTAR}
\def\xmm{XMM-Newton}

\def\xrt{{Swift}-XRT}

\def\maxi{MAXI}

\def\xrism{XRISM}
\def\xtend{XRISM-Xtend}

\def\hess{H.E.S.S.}

\newcommand{\HI}{H\,\textsc{i}}

\def\uG{\hbox{$\mu \mathrm{G}$}}

\def\eflux{\hbox{${\rm erg}~{\rm cm}^{-2}~{\rm s}^{-1}$}} 

\def\kms{\hbox{${\rm km}~ {\rm s}^{-1}$}}
\def\cc{\hbox{${\rm cm}^{-3}$}}
\def\column{\hbox{${\rm cm}^{-2}$}}  

\def\nt{n_2}

\def\Msun{\hbox{$M_{\odot}$}}

\def\NH{\hbox{$N_{\rm H}$}}

\usepackage{acro}

\DeclareAcronym{gde}{
  short = GDE ,
  long  = Galactic Diffuse Emission,
}

\DeclareAcronym{cgb}{
  short = CGB ,
  long  = cosmic gamma-ray background,
}

\DeclareAcronym{grb}{
  short = GRB ,
  long  = gamma-ray burst,
}
 
\DeclareAcronym{agn}{
  short = AGN ,
  long  = active galactic nucleus ,
  long-plural-form = active galactic nuclei
}
\DeclareAcronym{fsrq}{
  short = FSRQ ,
  long  = flat-spectrum radio quasar,
}
  
\DeclareAcronym{lmxb}{
  short = LMXB ,
  long  = low mass X-ray binary ,
  long-plural-form = low mass X-ray binaries ,
}
\DeclareAcronym{hmxb}{
  short = HMXB ,
  long  = high mass X-ray binary ,
  long-plural-form = high mass X-ray binaries ,
}
  
\DeclareAcronym{ic}{
  short = IC ,
  long  = inverse Compton ,
}

\DeclareAcronym{nustar}{
  short = {\it NuSTAR} ,
  long  = Nuclear Spectroscopic Telescope Array ,
}

\DeclareAcronym{bi}{
  short = BI ,
  long  = backside illumination ,
}
\DeclareAcronym{fi}{
  short = FI ,
  long  = frontside illumination ,
}
\DeclareAcronym{fov}{
  short = FoV ,
  long  = field of view ,
}
  
\DeclareAcronym{sim}{
  short = SIM ,
  long  = Science Instrument Module ,
}
\DeclareAcronym{hetg}{
  short = HETG ,
  long  = High Energy Transmission Grating ,
}
\DeclareAcronym{letg}{
  short = LETG ,
  long  = Low Energy Transmission Grating ,
}
\DeclareAcronym{hrc}{
  short = HRC ,
  long  = High Resolution Camera ,
}
\DeclareAcronym{acis}{
  short = ACIS ,
  long  = Advanced CCD Imaging Spectrometer ,
}
\DeclareAcronym{hrma}{
  short = HRMA ,
  long  = High Resolution Mirror Assembly,
}

\DeclareAcronym{compton}{
  short = Compton ,
  long  = Compton Gamma Ray Observatory,
}
\DeclareAcronym{hst}{
  short = HST ,
  long  = Hubble Space Telescope ,
}
\DeclareAcronym{iact}{
  short = IACT ,
  long  = Imaging Atmospheric Cherenkov Telescope ,
}
\DeclareAcronym{wcd}{
  short = WCD ,
  long  = Water Cherenkov Detector ,
}

\DeclareAcronym{hawc}{
  short = HAWC ,
  long  = High-Altitude Water Cherenkov ,
}
\DeclareAcronym{cta}{
  short = CTA ,
  long  = Cherenkov Telescope Array ,
}

\DeclareAcronym{em}{
  short = EM ,
  long  = electromagnetic ,
}
\DeclareAcronym{ism}{
  short = ISM ,
  long  = interstellar medium ,
}
\DeclareAcronym{csm}{
  short = CSM ,
  long  = circumstellar medium ,
}
\DeclareAcronym{sne}{
  short = SNe ,
  long  = supernovae , 
}

\DeclareAcronym{iss}{
  short = ISS ,
  long  = International Space Station ,
}

\DeclareAcronym{uhecr}{
  short = UHECRs ,
  long  = ultra high energy cosmic rays , 
}
\DeclareAcronym{ta}{
  short = TA ,
  long  = Telescope Array , 
}
\DeclareAcronym{auger}{
  short = Auger ,
  long  = Pierre Auger Observatory , 
}
\DeclareAcronym{ams}{
  short = AMS ,
  long  = Alpha Magnetic Spectrometer , 
}
\DeclareAcronym{pamela}{
  short = PAMELA ,
  long  = Payload for Antimatter Matter Exploration and Light-nuclei Astrophysics , 
}
   
\DeclareAcronym{cmb}{
  short = CMB ,
  long  = Cosmic Microwave Background , 
}
\DeclareAcronym{sed}{
  short = SED ,
  long  = spectral energy distribution , 
}
\DeclareAcronym{mhd}{
  short = MHD ,
  long  = magnetohydrodynamical ,
}
\DeclareAcronym{dof}{
  short = dof ,
  long  = degree of freedom ,
}
\DeclareAcronym{cco}{
  short = CCO ,
  long  = central compact object ,
  first-style = default
}
\DeclareAcronym{lmc}{
  short = LMC ,
  long  = Large Magellanic Cloud ,
}
\DeclareAcronym{smc}{
  short = SMC ,
  long  = Small Magellanic Cloud ,
}

\DeclareAcronym{hess}{
  short = H.E.S.S. ,
  long  = High Energy Stereoscopic System ,
  first-style = default
}
\DeclareAcronym{snr}{
  short = SNR ,
  long  = supernova remnant ,
}
\DeclareAcronym{pwn}{
  short = PWN ,
  short-plural = e ,
  long  = pulsar wind nebula ,
  long-plural  = e ,
}

\DeclareAcronym{sn}{
  short = SN ,
  short-plural = e ,
  long  = supernova ,
  long-plural  = e ,
  first-style = default
}

\DeclareAcronym{nw}{
  short = NW ,
  long  = northwest ,
  first-style = default
}
\DeclareAcronym{hxc}{
  short = HXC ,
  long  = hard X-ray component ,
  first-style = default
}

\DeclareAcronym{cr}{
  short = CR ,
  long  = cosmic ray ,
}
\DeclareAcronym{psf}{
  short = PSF ,
  long  = point spread function ,
}
\DeclareAcronym{hpd}{
  short = HPD ,
  long  = half power diameter ,
}
\DeclareAcronym{fwhm}{
  short = FWHM ,
  long  = full width of half maximum ,
}

\DeclareAcronym{pic}{
  short = PIC ,
  long  = particle-in-cell ,
  tag = numerical ,
}
\DeclareAcronym{cxb}{
  short = CXB ,
  long  = Cosmic X-ray Background ,
}
\DeclareAcronym{grxe}{
  short = GRXE ,
  long  = Galactic Ridge X-ray Emission ,
}
\DeclareAcronym{pa}{
  short = PA ,
  long  = Positional Angle ,
}
\DeclareAcronym{dsa}{
  short = DSA ,
  long  = diffusive shock acceleration ,
}

\def\nt{n_2}

\def\NH{\hbox{$N_{\rm H}$}}

\def\degr{\hbox{$^\circ$}}
\def\arcmin{\hbox{$^\prime$}}

\def\utw{\smash{\rlap{\lower5pt\hbox{$\sim$}}}}
\def\udtw{\smash{\rlap{\lower6pt\hbox{$\approx$}}}}
\def\fd{\hbox{$.\!\!^{\rm d}$}}
\def\fh{\hbox{$.\!\!^{\rm h}$}}
\def\fm{\hbox{$.\!\!^{\rm m}$}}
\def\fs{\hbox{$.\!\!^{\rm s}$}}

\def\apjl{ApJ}%
\def\apjs{ApJS}%
\def\ao{Appl.~Opt.}%
\def\aap{A\&A}%
\usepackage{physics}
\usepackage{upgreek}
\usepackage{CJK}

\def\nt#1{\textcolor{magenta}{[NT] #1}} %
\def\revise#1{\textbf{#1}} %

\begin{document}
\begin{CJK*}{UTF8}{gbsn}

\title{
XMM-Newton Resolves Parsec-scale X-ray Jets in the PeVatron Microquasar V4641 Sgr
}

\author[0000-0001-7209-9204]{Naomi Tsuji}
\affiliation{Institute for Cosmic Ray Research, The University of Tokyo, 5-1-5, Kashiwa-no-ha, Kashiwa, Chiba 277-8582, Japan}
\affiliation{Interdisciplinary Theoretical \& Mathematical Science Center (iTHEMS), RIKEN, 2-1, Hirosawa, Wako, Saitama 351-0198, Japan}
\email[show]{ntsuji@icrr.u-tokyo.ac.jp}
\author[0009-0002-1992-0707]{Caterina Tresoldi}
\affiliation{INAF-Osservatorio Astronomico di Brera, Via E. Bianchi 46, 23807 Merate (LC), Italy}
\affiliation{Como Lake Center for Astrophysics (CLAP), DiSAT, Università degli Studi dell’Insubria, via Valleggio 11, 22100 Como, Italy}
\email[show]{caterina.tresoldi@inaf.it}
\author[0000-0002-9709-5389]{Kaya Mori}
\affiliation{Columbia Astrophysics Laboratory, Columbia University, 538 West 120th Street, New York, NY 10027, USA}
\email{kaya@astro.columbia.edu}
\author[0000-0002-8152-6172]{Hiromasa Suzuki}
\affiliation{The University of Miyazaki, Miyazaki, 1-1 Gakuen-Kibanadai-Nishi, Miyazaki, 889-2192 Japan}
\email{suzuki@astro.miyazaki-u.ac.jp}
\author[0000-0002-9709-5389]{Matthew Lundy}
\affiliation{Columbia Astrophysics Laboratory, Columbia University, 538 West 120th Street, New York, NY 10027, USA}
\email{mwl2145@columbia.edu}
\author[0000-0002-7576-7869]{Dmitry Khangulyan}
\affiliation{Key Laboratory of Particle Astrophysics, Institute of High Energy Physics, Chinese Academy of Sciences, Beijing 100049, China}
\affiliation{Tianfu Cosmic Ray Research Center, 610000 Chengdu, Sichuan, China}
\email{khangulyan@ihep.ac.cn}
\author[0000-0001-6189-7665]{Samar Safi-Harb}
\affiliation{Department of Physics and Astronomy, University of Manitoba, Winnipeg, MB R3T 2N2, Canada}
\email{Samar.Safi-Harb@umanitoba.ca}
\author[0000-0002-9105-0518]{Laura Olivera-Nieto}
\affiliation{Anton Pannekoek Institute for Astronomy, University of Amsterdam, Science Park 904, 1098 XH Amsterdam, The Netherlands}
\affiliation{Max-Planck-Institut f\"{u}r Kernphysik, Saupfercheckweg 1, 69117 Heidelberg, Germany}
\email{laura.olivera-nieto@mpi-hd.mpg.de}
\author[0000-0003-1157-3915]{Felix Aharonian}
\affiliation{Yerevan State University, 1 Alek Manukyan Street, Yerevan 0025, Armenia}
\affiliation{Max-Planck-Institut f\"{u}r Kernphysik, Saupfercheckweg 1, 69117 Heidelberg, Germany}
\email{felix.aharonian@mpi-hd.mpg.de}
\author[0000-0001-7250-3596]{Shotaro Abe}
\affiliation{Department of Physics, Kyoto University, Kitashirakawa Oiwake-cho, Sakyo, Kyoto, 606-8502, Japan}
\email{shotaro@icrr.u-tokyo.ac.jp}
\author[0000-0002-4858-6130]{Zhen Cao}
\affiliation{Key Laboratory of Particle Astrophysics, Institute of High Energy Physics, Chinese Academy of Sciences, Beijing 100049, China}
\affiliation{Tianfu Cosmic Ray Research Center, 610000 Chengdu, Sichuan, China}
\affiliation{University of Chinese Academy of Sciences, 100049 Beijing, China}
\email{caozh@ihep.ac.cn}
\author[0000-0002-6144-9122]{Sabrina Casanova}
\affiliation{Institute of Nuclear Physics, ul. Radzikowskiego 152, 31-342 Krak\'{o}w, Poland}
\email{sabrina.casanova@ifj.edu.pl}
\author[0000-0002-9735-3608]{Maria Chernyakova}
\affiliation{School of Physical Sciences and Centre for Astrophysics \& Relativity, Dublin City University, Glasnevin, D09 W6Y4, Ireland}
\email{masha.chernyakova@dcu.ie}
\author[0000-0002-2421-173X]{Yi-Heng Chi}
\affiliation{School of Astronomy \& Space Science, Nanjing University, 163 Xianlin Avenue, Nanjing 210023, China}
\email{yhchi@smail.nju.edu.cn}
\author[0000-0001-5538-5831]{St\'ephane Corbel}
\affiliation{Universit\'e Paris Cit\'e, Universit\'e Paris-Saclay, CEA, CNRS, AIM, F-91191, Gif-sur-Yvette, France}
\email{stephane.corbel@cea.fr}
\author[0000-0001-9643-4134]{E. de la Fuente}
\affiliation{Departamento de F\'{i}sica, CUCEI, Universidad de Guadalajara, Blvd. Marcelino Garc\'{i}a Barragan 1420, Ol\'{i}mpica, 44430, Guadalajara, Jalisco, M\'exico}
\email{eduardo.delafuente@academicos.udg.mx}
\author[0000-0001-7584-6236]{Hua Feng}
\affiliation{Key Laboratory of Particle Astrophysics, Institute of High Energy Physics, Chinese Academy of Sciences, Beijing 100049, China}
\email{hfeng@ihep.ac.cn}
\author[0009-0002-9370-3313]{Tatsuki Fujiwara}
\affiliation{Department of Earth and Space Science, Graduate School of Science, The University of Osaka, 1-1 Machikaneyama, Toyonaka, Osaka 560-0043, Japan}
\email{u000719k@ecs.osaka-u.ac.jp}
\author[0009-0005-4358-5146]{Noa Grollimund}
\affiliation{Universit\'e Paris Cit\'e, Universit\'e Paris-Saclay, CEA, CNRS, AIM, F-91191, Gif-sur-Yvette, France}
\email{noa.grollimund@u-pariscite.fr}
\author[0000-0002-5447-1786]{Dezhi Huang}
\affiliation{Department of Physics and Astronomy, University of Delaware, Newark, DE, USA}
\email{dezhih@mtu.edu}
\author[0000-0002-7272-1136]{Yoshiyuki Inoue}
\affiliation{College of Systems Engineering and Science, Shibaura Institute of Technology, 307 Fukasaku, Minuma-ku, Saitama City, Saitama
337-8570, Japan}
\affiliation{Department of Earth and Space Science, Graduate School of Science, The University of Osaka, 1-1 Machikaneyama, Toyonaka, Osaka 560-0043, Japan}
\affiliation{Interdisciplinary Theoretical \& Mathematical Science Center (iTHEMS), RIKEN, 2-1, Hirosawa, Wako, Saitama 351-0198, Japan}
\affiliation{Kavli Institute for the Physics and Mathematics of the Universe (WPI), The University of Tokyo, Kashiwa 277-8583, Japan}
\email{yinoue@astro-osaka.jp}
\author[]{Ryohei Kanazashi}
\affiliation{Faculty of Science, Kanagawa University, 3-27-1 Rokukakubashi, Kanagawa-ku, Yokohama, Kanagawa 221-8686}
\email{r202570038oq@jindai.jp}
\author[orcid=0000-0003-2579-7266]{Shigeo S. Kimura}
\affiliation{Astronomical Institute, Graduate School of Science, Tohoku University, Sendai 980-8578, Japan}
\affiliation{Frontier Research Institute for Interdisciplinary Sciences, Tohoku University, Sendai 980-8578, Japan}
\email{shigeo@astr.tohoku.ac.jp}
\author[0000-0002-4462-3686]{Marianne Lemoine-Goumard}
\affiliation{Universit\'{e} Bordeaux, CNRS, CENBG, UMR 5797, F-33170 Gradignan, France}
\email{lemoine@cenbg.in2p3.fr}
\author[0000-0002-5449-6131]{Jonathan Mackey}
\affiliation{I.~Physikalisches Institut, Universit\"{a}t zu K\"{o}ln, Z\"{u}lpicher Str. 77, 50937 K\"{o}ln, Germany}
\affiliation{Astronomy \& Astrophysics Section, School of Cosmic Physics, Dublin Institute for Advanced Studies, DIAS Dunsink Observatory, Dublin D15 XR2R, Ireland}
\affiliation{School of Physics, University College Dublin, Belfield, Dublin 4, Ireland}
\email{jmackey@cp.dias.ie}
\author[0000-0001-5674-4664]{Alessio Marino}
\affiliation{Institute of Space Sciences (ICE, CSIC), Campus UAB, Carrer de Can Magrans s/n, 08193, Barcelona, Spain}
\affiliation{Institut d'Estudis Espacials de Catalunya (IEEC), Esteve Terradas 1, RDIT Building, Of. 212 Mediterranean Technology Park (PMT), 08860, Castelldefels, Spain}
\affiliation{Departamento de F\'{i}sica, Universidad de Santiago de Chile (USACH), Av. V\'{i}ctor Jara 3493, Estaci\'{o}n Central, Chile}
\affiliation{Center for Interdisciplinary Research in Astrophysics and Space Sciences (CIRAS), Universidad de Santiago de Chile}
\affiliation{INAF/IASF Palermo, via Ugo La Malfa 153, I-90146 - Palermo, Italy}
\email{marino@ice.csic.es}
\author[]{Samaresh Mondal}
\affiliation{Department of Astronomy, University of Illinois, 1002 W. Green St., Urbana, IL 61801, USA}
\email{mondalsamaresh088@gmail.com}
\author[0009-0004-5641-5552]{Kai Morita}
\affiliation{Institute for Cosmic Ray Research, The University of Tokyo, 5-1-5, Kashiwa-no-ha, Kashiwa, Chiba 277-8582, Japan}
\email{morikai@icrr.u-tokyo.ac.jp}
\author[0009-0003-8610-853X]{Maxime Parra}
\affiliation{Department of Physics, Ehime University, 2-5 Bunkyocho, Matsuyama, Ehime 790-8577, Japan}
\email{maximeparrastro@gmail.com}
\author[0000-0003-0293-3608]{Gabriele Ponti}
\affiliation{INAF-Osservatorio Astronomico di Brera, Via E. Bianchi 46, 23807 Merate (LC), Italy}
\affiliation{Como Lake Center for Astrophysics (CLAP), DiSAT, Università degli Studi dell’Insubria, via Valleggio 11, 22100 Como, Italy}
\email{gabriele.ponti@inaf.it}
\author[0000-0002-3778-1432]{Brian Reville}
\affiliation{Max-Planck-Institut f\"{u}r Kernphysik, Saupfercheckweg 1, 69117 Heidelberg, Germany}
\email{brian.reville@mpi-hd.mpg.de}
\author[0000-0001-8195-6546]{Megumi Shidatsu}
\affiliation{Department of Physics, Ehime University, 2-5 Bunkyocho, Matsuyama, Ehime 790-8577, Japan}
\email{shidatsu.megumi.wr@ehime-u.ac.jp}
\author[0000-0001-8147-6817]{Shunya Takekawa}
\affiliation{Faculty of Engineering, Kanagawa University, 3-27-1 Rokkakubashi, Kanagawa-ku, Yokohama, Kanagawa 221-8686, Japan}
\email{shunya@kanagawa-u.ac.jp}
\author[0000-0003-3417-1425]{Michelle Tsirou}
\affiliation{Institute of Physics of the Czech Academy of Sciences, Prague, 182 00, Czechia}
\email{tsirou@fzu.cz}
\author[0000-0001-6798-353X]{Xiaojie Wang}
\affiliation{Department of Physics, Missouri University of Science and Technology, Rolla, MO, USA}
\email{xiaojiewang@mst.edu}
\author[0000-0003-2840-6152]{Heshou Zhang}
\affiliation{INAF-Osservatorio Astronomico di Brera, Via E. Bianchi 46, 23807 Merate (LC), Italy}
\email{heshou.zhang@inaf.it}
\author[0000-0002-5683-822X]{Ping Zhou}
\affiliation{School of Astronomy \& Space Science, Nanjing University, 163 Xianlin Avenue, Nanjing 210023, China}
\affiliation{Key Laboratory of Modern Astronomy and Astrophysics, Nanjing University, Ministry of Education, Nanjing 210023, China}
\email{pingzhou@nju.edu.cn}


\correspondingauthor{Naomi Tsuji, Caterina Tresoldi}


\begin{abstract}
The microquasar V4641 Sagittarii (V4641 Sgr) has recently been established as one of the most powerful Galactic PeV cosmic-ray accelerators (PeVatrons) through the detection of degree-scale gamma-ray emission to 0.8 PeV. Using XMM-Newton, we performed an X-ray survey of its gamma-ray-emitting regions during the quiescent phase.
We discovered extended X-ray emission to the north and south of V4641 Sgr, aligned with the radio-jet axis, revealing X-ray jet structures extending over several tens of parsecs for the first time.
The northern and southern knot-like features, dubbed n1 and s1, are located approximately 20 pc from the central binary and spatially coincident with the gamma-ray emission.
The X-ray emission from each knot, characterized by a photon index of $\Gamma_X = 1.9 \pm 0.1_{\rm stat} \pm 0.5_{\rm syst}$ and an energy flux of $\sim$(2--4)$\times10^{-13}$~erg~cm$^{-2}$~s$^{-1}$ in the 2--10 keV band, is likely non-thermal in origin, indicating synchrotron radiation from accelerated electrons.
The morphological coincidence between the X-ray and gamma-ray emission favors a leptonic scenario for the gamma rays, although the X-ray emission is more compact than the gamma-ray-emitting region and the hadronic scenario cannot be excluded.
Through X-ray to gamma-ray modeling, we obtain a weak magnetic field strength of 3--5~$\mu$G.
We propose that the newly detected X-ray knots represent particle-acceleration sites, such as jet termination or (re-)collimation regions powered by V4641 Sgr, or downstream structures analogous to the knots of the outer jets in SS~433/W50.
This discovery suggests that extended X-ray emission along jets would be a key signature, yet previously overlooked, in PeVatron microquasars.

\end{abstract}



\section{Introduction}

Recent advances in air-shower arrays, such as Tibet AS$\gamma$, HAWC, and LHAASO, have opened up a new era in PeV gamma-ray astronomy.
In particular, the detection of very-high-energy (VHE; $E>100$~GeV) and ultra-high-energy (UHE; $E>100$~TeV) gamma rays from microquasars --- X-ray binary systems,
containing a compact object (black hole or neutron star) with a companion star and launching relativistic jets
--- has established them as a new class of PeV cosmic-ray accelerators, namely PeVatrons.
SS~433 and its surrounding nebula W50 represent the first identified PeVatron microquasar \citep{hawc_2018_ss433}.
The second example is V4641 Sagittarii \citep[V4641 Sgr; ][]{hawc_2024_v4641}, followed by detections of three additional microquasars, GRS~1915$+$105, MAXI~J1820$+$070, and Cygnus~X-1, in the VHE-UHE gamma-ray band \citep{lhaaso_2025_microquasars}.
Furthermore, another microquasar, Cygnus X-3, has recently been identified as a hadronic super-PeVatron, namely a particle accelerator capable of exceeding multiple PeV energy ranges
\citep{lhaasocollaborationCygnusX3Variable2025}.
Particle acceleration in microquasars has therefore attracted increasing attention.
In addition to currently ``active" microquasars being considered promising PeVatron candidates, so-called microquasar ``remnants" --- systems that were active in the past but are no longer producing jets or outbursts --- have also been proposed as counterparts of VHE-UHE unidentified gamma-ray sources \citep{abaroaMicroquasarRemnantsHidden2025,Kaci_2025,Zhang_2026}.
We note that nearly half of the sources in the first LHAASO catalog are presently unidentified sources \citep{cao_first_2024}.

The \src\ system is the second microquasar identified as a PeVatron and the most luminous TeV gamma-ray-emitting microquasar.
Its gamma-ray emission has been detected up to 0.8~PeV, exhibiting a spatial extension on a scale of $\sim1$\degr\ and one of the hardest spectra among known Galactic TeV gamma-ray sources in $\sim$1--10 TeV with a photon index of $\Gamma_\gamma \approx 1.8$ \citep{hawc_2024_v4641,lhaaso_2025_microquasars,hess_v4641_2026}.
The central object is a low-mass X-ray binary consisting of a 6.4 \Msun\ black hole and a 2.9 \Msun\ B9III companion star, located at a distance of $d \approx 6.2$~kpc \citep{MacDonald_2014_v4641,gaia_2019_BHs}.
\src\ underwent a major super-Eddington outburst in 1999 and has subsequently exhibited recurrent outbursts roughly every two years \citep[e.g.,][]{Revnivtsev_2002_v4641}.
Radio jets were confirmed during the 1999 X-ray outburst by observations with the Very Large Array \citep{1999IAUC.7119....1I,Hjellming_2000_v4641}, and recent re-analysis suggested that the jet was launched nearly perpendicular to the line of sight, with its direction aligned with the extent of the TeV gamma-ray emission \citep{Marti_2026_v4641}.
Moreover, deep 1.3~GHz radio continuum observations with MeerKAT revealed a so-called ``bow-tie" structure, likely powered by the jet or wind from \src\ \citep{Grollimund_2026_v4641_meerkat}.

In the X-ray band, as with other X-ray binaries, observations of \src\ have primarily focused on its outburst phases. Consequently, the degree-scale gamma-ray emission has not been fully covered by archival X-ray data.
Approximately half of the gamma-ray-emitting region was observed with the wide-field CCD camera Xtend onboard \xrism, leading to the first detection of extended X-ray emission likely associated with the gamma-ray radiation \citep{suzukiDetectionExtendedXRay2025}. However, this \xrism\ observation was conducted immediately after the outburst in 2024 September \citep{Parra_2025_v4641}, resulting in contamination from the bright central source. In addition, the 38\arcmin\ field of view (FoV) of Xtend did not fully cover the entire gamma-ray emission region.
Recently, \cite{Bykov_2026_v4641} presented an SRG/eROSITA map around \src, placing X-ray flux upper limits on the entire gamma-ray-emitting region.
Therefore, the X-ray counterpart of the degree-scale TeV emission remains unexplored.
Nevertheless, the importance of X-ray observations of TeV gamma-ray-emitting microquasars has already been demonstrated through extensive studies of the SS~433/W50 system \citep[e.g.,][]{safi-harbHardXRayEmission2022a,kayamaSpatiallyResolvedStudy2022,kayamaXrayStudyPropagation2025,chiXRayShellReveals2024,kaaretXRayPolarizationEastern2024,tsujiFirstProperMotion2025,Sunyaev_2026_ss433,MacIntyre_2026_ss433}.
Non-thermal X-ray emission traces multi-TeV electrons through synchrotron radiation and provides strong constraints on the magnetic-field strength in leptonic scenarios for the gamma-ray emission. Furthermore, the high angular resolution of X-ray satellites enables the identification and detailed characterization of particle-acceleration sites within microquasar systems.

In this Letter, we report the first, deep X-ray survey of the gamma-ray emitting region surrounding \src\ during its quiescent phase by \xmm, resulting in the discovery of two extended knot-like features within the gamma-ray emission.
This work is coordinated with a companion study of X-ray observations with Einstein Probe \citep{Chi_2026_EP}.
Combined with the scheduled and ongoing follow-up observations with \chandra\ and \nustar, a comprehensive broadband X-ray study will be presented in separate publication(s).

\section{Observations and data reduction} \label{sec:obs}

We performed \xmm\ observations of the region surrounding V4641 Sgr on September 2025, covering the entire VHE-UHE gamma-ray emission (Table~\ref{table:obs} in Appendix~\ref{appendix: X-ray data set}).
The observations consist of four pointings with 30--40~ks exposure each, except soft proton flares reduced the effective time to $\sim$10 ks for one observation (ObsID: 0971600401).

It should be noted that the \xmm\ observations were done during the quiescent phase of \src.
Prior to the observations, we monitored the activity of the central black hole with \maxi, \xrt, and Einstein Probe.
In particular, we conducted a \xrt\ DDT observation with PC mode in September 6 with effective exposure of 772~s (ObsID: 00030111115), yielding 0.3--10~keV flux upper limits of $2.5 \times 10^{-13}$~\eflux\ at 90\% confidence level (c.l.) with a column density (\NH) and photon index ($\Gamma_X$) being fixed to $1.5 \times 10^{21}$~cm$^{-2}$ and 1.5, respectively.
The flux upper limit of the central black hole was $1.5 \times10^{-14}$~\eflux\ during our \xmm\ observations on September 13.
This flux was approximately three orders of magnitude lower than that measured during the \xrism\ observation \citep[i.e., $\sim 3 \times 10^{-11}$~\eflux;][]{suzukiDetectionExtendedXRay2025,Parra_2025_v4641}, which significantly reduced the contamination from the central object.

The data were reduced and analyzed using the Science Analysis System (SAS) v22.1.0\footnote{\url{https://www.cosmos.esa.int/web/xmm-newton/sas-release-notes-2210}}. We made use of data from all three European Photon Imaging Cameras (EPIC), namely MOS1, MOS2, and PN, operating in Full Frame mode.
We followed the data reduction procedure described in detail by \citet{2012A&A...545A.128H} and \citet{2015MNRAS.453..172P}.
Calibrated event files were produced using the SAS tasks \texttt{emchain} and \texttt{epchain} for the MOS and PN cameras, respectively.
For the PN detector, we also generated Out-of-Time (OoT) event files (\texttt{withoutoftime=Y}) to account for the OoT effect and ensure a proper correction for the Charge Transfer Inefficiency (CTI).
In order to filter soft proton flares, we first generated Good Time Interval (GTI) files starting from the 7--15 keV background light curves using the \texttt{tabgtigen} task, applying a conservative count-rate threshold of 2.5~counts~s$^{-1}$ for the MOS and 8.0~counts~s$^{-1}$ for the PN detectors. Subsequently, we refined the flare filtering using the histogram-based method implemented in the \texttt{espfilt} task within the \xmm\ Extended Source Analysis Software (XMM-ESAS) package\footnote{\url{https://www.cosmos.esa.int/web/xmm-newton/xmm-esas}}. This filtering was performed directly on the observation event list over the standard 2.5–8 keV energy band, and the task identifies flare-free intervals by fitting a Gaussian distribution to the count-rate histogram and selecting the time intervals within 2.5$\sigma$.
The effective exposure times after the soft-proton cleaning procedure are reported in Appendix~\ref{appendix: X-ray data set}.

\section{Analysis and Results}

Using the filtered event lists (Section~\ref{sec:obs}), we produced mosaic images and extracted source spectra, as described in Section~\ref{sec:image} and Section~\ref{sec:spectra}, respectively.

\subsection{X-ray Imaging}
\label{sec:image}

\begin{figure}[tb!]
\centering
\includegraphics[width=\linewidth]{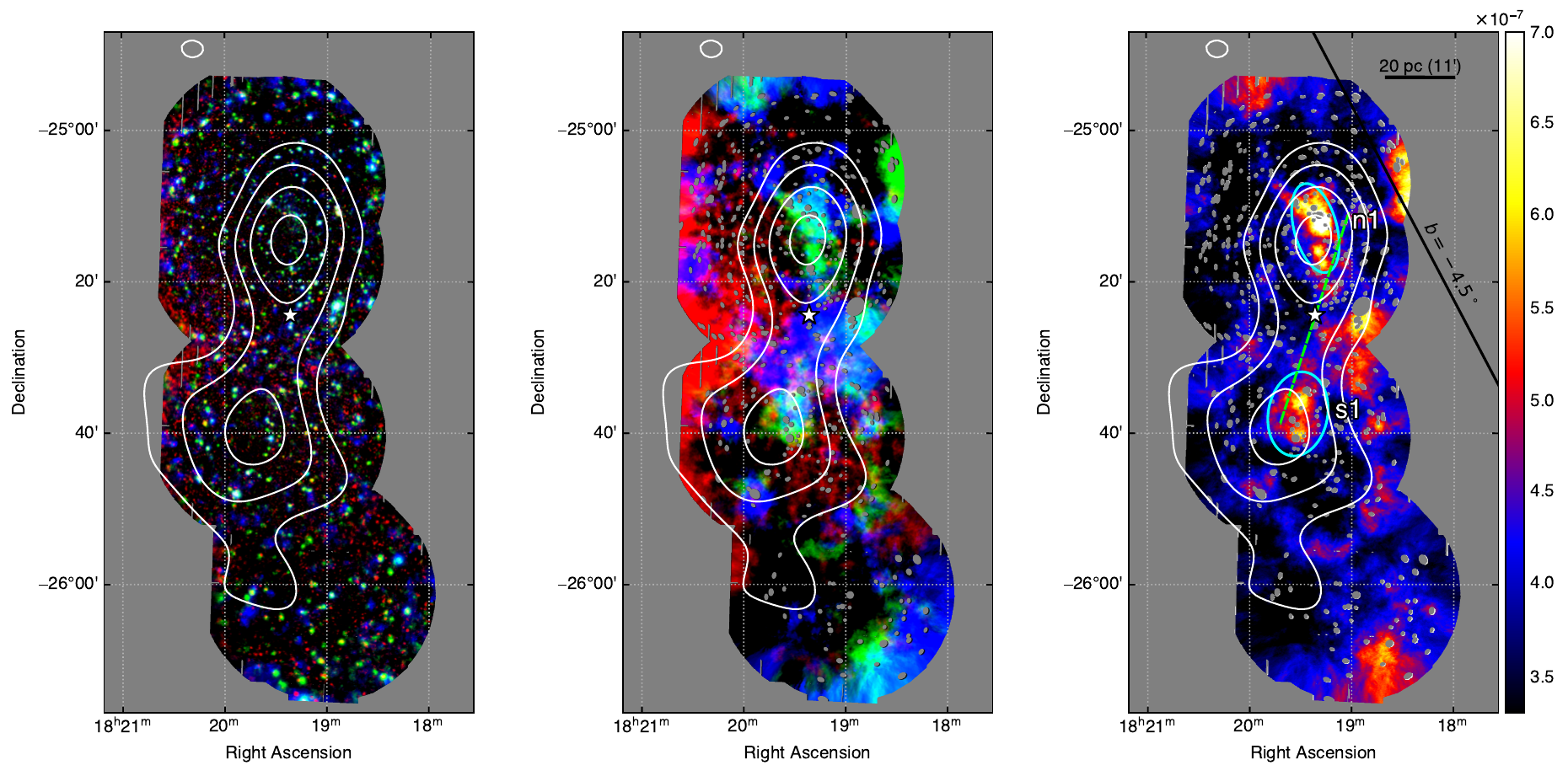}
  \caption{
  \textit{Left}:
  The RGB images obtained by \xmm\ (0.5--2~keV in red, 2--5~keV in green, and 5--8~keV in blue).
  The star denotes the position of \src.
  The images are background-subtracted, exposure-corrected for vignetting, and adaptively smoothed with a minimum signal-to-noise ratio of 3.
  All three EPIC instruments and all observation epochs are combined.
  The $>$0.8 TeV gamma-ray emission observed by \hess\footnote{\url{https://zenodo.org/records/17570420}} (3, 5, 7, and 9$\sigma$ significance levels) is shown with the white contours.
  \textit{Middle}:
  Same as the left panel, but point-like sources are masked and smoothed with a minimum signal-to-noise ratio of 15.
    \textit{Right}: The \xmm\ image in 2--5 keV, shown with newly detected extended X-ray emission, n1 and s1.
    The cyan ellipses show the spectral extraction regions.
    The dashed line is the radio jet axis \citep{Marti_2026_v4641}.
    Fo all panels, the map is in units of counts~s$^{-1}$ and the pixel size is 2$^{\prime\prime}$.
\label{fig:map}}
\end{figure}

Count maps and exposure maps were generated with the \texttt{eimageget} task.
The detector background derived from filter wheel closed data was subtracted for each observation, following \cite{2016A&A...585A.162M}.
The images were then exposure-corrected.
Noisy CCDs in the MOS data \citep{2008A&A...478..575K} were identified using the SAS task \texttt{emtaglenoise} and excluded from the images.
For each detector and energy band, background-subtracted mosaics were produced by combining all four observations.
MOS and PN maps were subsequently merged into a single count-rate map, applying an exposure cut-off of 7.2 ks. The resulting mosaic was then adaptively smoothed using the \texttt{exposure\_smooth} procedure in \texttt{contbin} package \citep{2006MNRAS.371..829S}.

The obtained images are illustrated in Figure~\ref{fig:map}, which show remarkably many point-like sources.
Since \src\ is located at ($\ell$, $b$) = (6.774\degr , $-$4.789\degr), these point-like sources consist of both Galactic sources \citep[i.e., coronally active stars and accreting white dwarfs;][]{Revnivtsev_2006,Revnivtsev_2009} and active galactic nuclei (AGNs). The detailed study of point sources will be carried out by the scheduled \chandra\ observations.
To search for extended diffuse X-ray emission, these point-like sources were removed as follows.
First, source detection was performed on each observation using a maximum-likelihood technique implemented in the SAS meta-task \texttt{edetect\_chain}\footnote{\url{https://xmm-tools.cosmos.esa.int/external/sas/current/doc/edetect_chain/}}, applied to images from the three EPIC instruments in the four energy bands (0.5--1.0~keV, 1.0--2.0~keV, 2.0--4.5~keV, and 4.5--12.0~keV).
Second, to accurately mask point sources, we used the SAS task \texttt{region} to define exclusion regions for all detected point-like sources. A contour radius style was adopted, which defines exclusion boundaries based on the point spread function (PSF) wings. A background fraction threshold of \texttt{bkgfraction=0.1} was applied, masking the source extent until its contribution falls to 10\% of the local background level.
This procedure effectively removes point-source contamination while preserving the largest possible area for the analysis of diffuse emission.

Figure~\ref{fig:map} (middle) shows the \xmm\ images after masking all point-like sources.
In the soft ($\lesssim 2$~keV) X-ray band, a largely extended emission was detected. However, this soft emission is likely unrelated to the gamma-ray emission, due to the lack of spatial correlation. Instead it is most likely related to Galactic foreground because of the plausible spatial anti-correlation between \HI\ and X-rays \citep{2016A&A...594A.116H,hawc_2024_v4641}.
The medium-band image (2--5~keV) revealed two promising extended X-ray features located near the gamma-ray centroids in the northern and southern regions. Following the notation of the known X-ray knots in SS~433 \citep[e.g.,][]{safi-harbROSATASCAObservations1997}, we designate these features as ``n1" and ``s1" (Figure~\ref{fig:map}; right).
In addition, marginal excesses were hinted at on the western side of \src\ and near the southern edge of the FoV.
The 2--5 keV image is compared with radio to gamma-ray images in Figure~\ref{fig:MWL}.
The n1 and s1 emissions are likely different components from the MeerKAT bow tie and not spatially coincident with any of molecular clouds (Figure~\ref{fig:MWL}b).
Note several $^{12}$CO emission presented in Figure~\ref{fig:MWL}b are located at $<4$~kpc or $>12$~kpc (traditional kinematic distances inferred from the measured velocity of molecular clouds and a Galactic rotation curve \cite[e.g.,][]{wenger_kinematic_2018}), meaning they are unlikely at the same distance as \src.
No clear hard X-ray emission was detected above 5~keV, except for a possible counterpart associated with the n1 region.
We note that the extended emission around the edges of FoVs may be artifacts and/or residual contamination from (unresolved) point sources.
No known source is spatially correlated with these extended X-ray diffuse emission, as described in \cite{hawc_2024_v4641}.

The locations of the newly discovered diffuse emission (n1 and s1) are summarized in Table~\ref{tab:fit}.
Table~\ref{tab:fit} also includes their spatial extent, such as major and minor axes and angle, fitted by a 2D Gaussian model with all parameters free.
Notably, n1 was independently covered in two separate observations (ObsIDs 0971600101 and 0971600201), confirming the presence in both epochs.
Both n1 and s1 are located approximately 12\arcmin\ (corresponding to 23 pc for $d=6.2$~kpc) from the central source and roughly aligned with the arcsecond-scale radio jets detected by VLA \citep{Marti_2026_v4641}, suggesting a possible association with the jets.
We note, however, that the directions from \src\ to n1 and s1 differ from the extrapolated radio-jet axis by $\sim16$\degr\ and $\sim10$\degr, respectively (Figure~\ref{fig:map}).
n1 and s1 are included within the 68\% gamma-ray containment region, supporting a spatial coincidence.
Whereas a two-component morphology in the TeV gamma-ray band was only slightly favored by \hess\ at the $\sim 3 \sigma$ level \citep{hess_v4641_2026}, n1 and s1 are clearly separated in the X-ray image.
Furthermore, this two lobe-like structure would exclude the scenario of dust scattering halo, discussed in \cite{suzukiDetectionExtendedXRay2025}.
%
The additional marginal excess of the western and southern components is centered at
(R.A., Decl.) = (18\fh18\fm56\fs64, $-$25\fd26\fm38\fs4) and (18\fh18\fm45\fs36, $-$26\fd06\fm39\fs6),
respectively. The latter emission has a bright radio source, PMN~J1818$-$2608, near its center; however, it is likely a background AGN and therefore unrelated to the diffuse X-ray emission.
The scheduled \chandra\ and \nustar\ observations will provide further constraints on the physical properties and existence of these new detected X-ray sources.


Here we compare our findings with \xmm\ with the previous XRISM results.
Figure~\ref{fig:MWL}c shows the images of \xmm\ (before masking out the point-like sources) and \xtend\ \citep{suzukiDetectionExtendedXRay2025}.
Although the point-source-removed \xmm\ image (Figure~\ref{fig:map}) looks different from the XRISM image, the diffuse emission reported by \xtend\ is roughly reproduced by a combination of n1, s1, and unresolved point-like sources, which the short-exposure \xrism\ observation was not sensitive enough to detect.

\begin{figure*}[tb!]
  \centering
    \begin{minipage}{\linewidth}
    \centering
    \includegraphics[width=\linewidth]{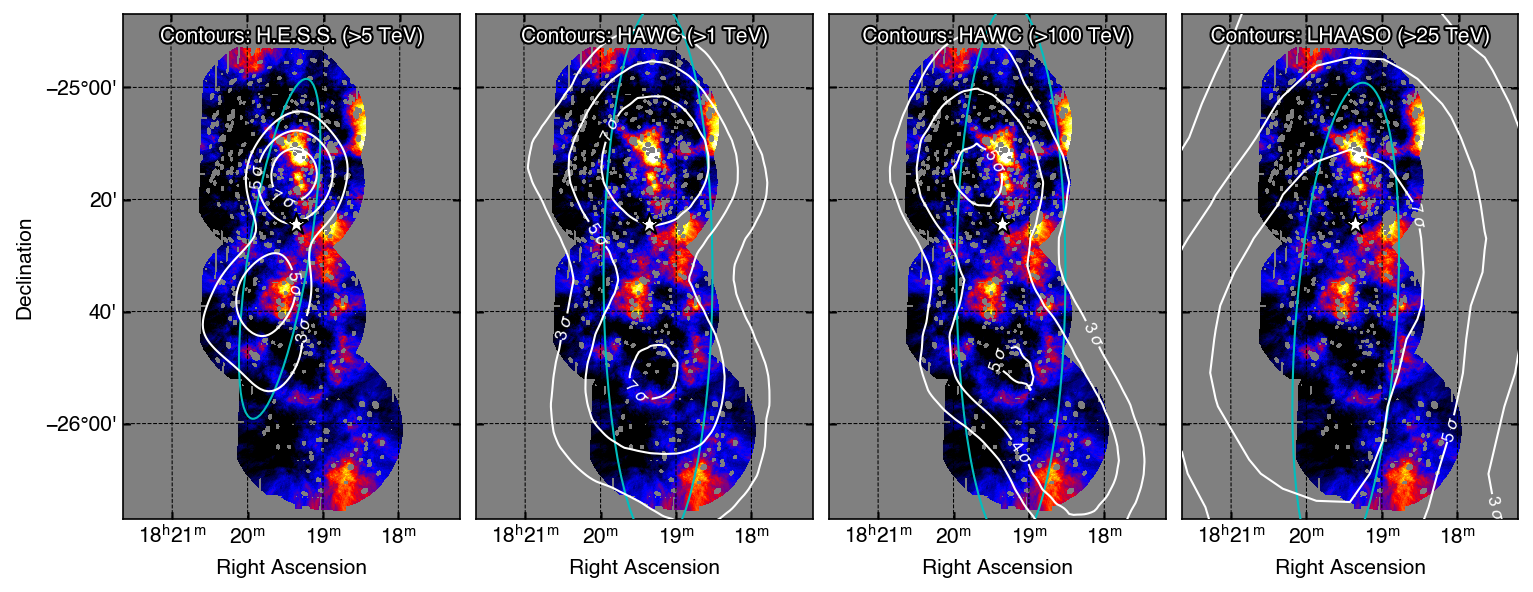}
    \subcaption{\xmm\ vs gamma rays.}
  \end{minipage}
  \\
  \begin{minipage}{\linewidth}
    \centering
    \includegraphics[width=\linewidth]{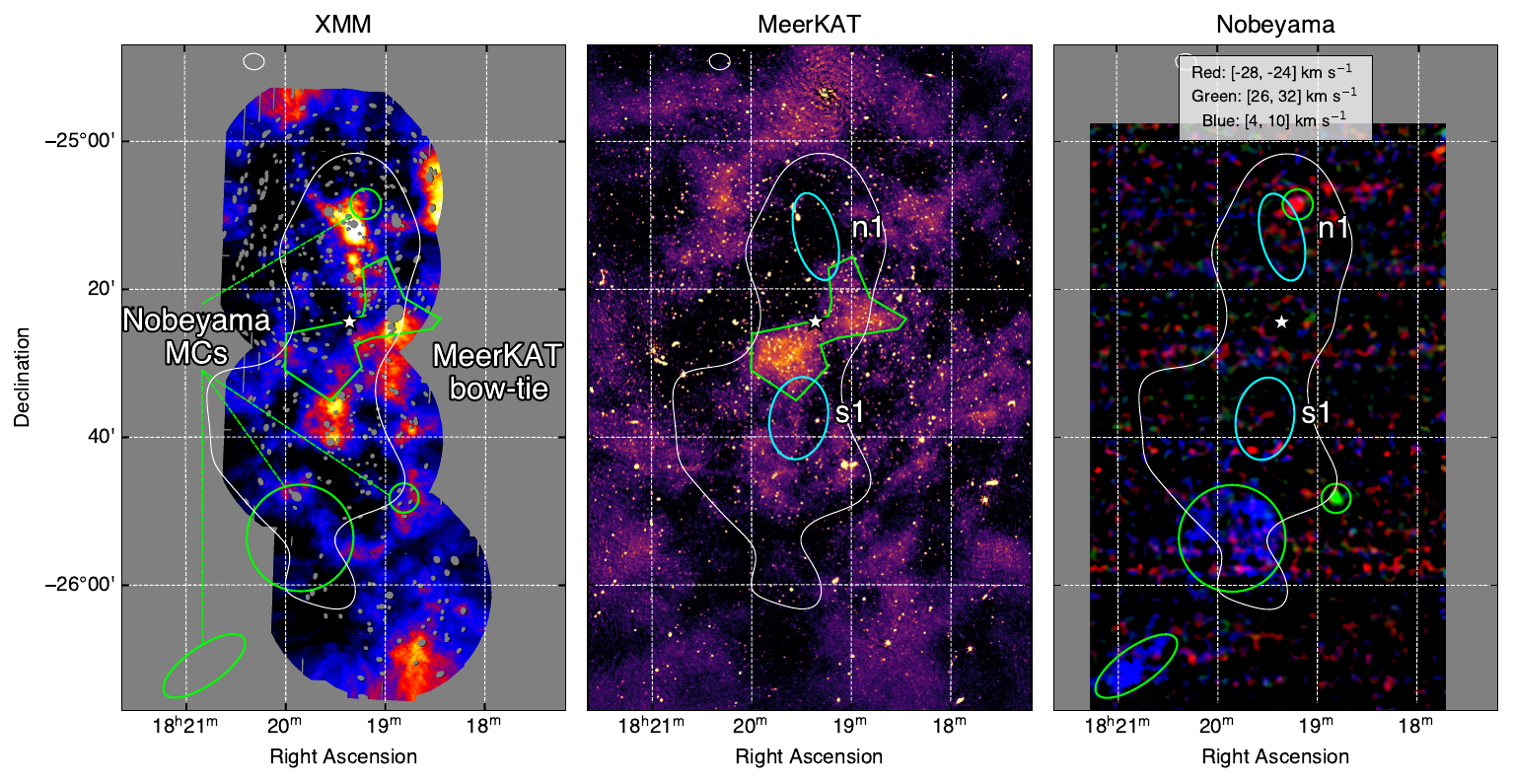}
    \subcaption{\xmm\ vs radio.}
  \end{minipage}
  %
  \caption{
Multi-wavelength view in the \src\ region.
(a) The \xmm\ 2--5 keV image, overlaid with contours of significance maps measured by \hess, HAWC\footnote{\url{https://data.hawc-observatory.org/datasets/V4641/index.php}}, and LHAASO\footnote{\url{https://english.ihep.cas.cn/lhaaso/pdl/202110/t20211026_286779.html}} \citep{hess_v4641_2026,hawc_2024_v4641,lhaaso_2025_microquasars}.
    Note that the PSF sizes are large in the gamma-ray instruments ($\gtrsim 0.1$\degr), compared to that of \xmm\ (6$^{\prime\prime}$ in FWHM).
    The cyan ellipse indicates the best-fit asymmetry 2D Gaussian model of the gamma-ray data.
(b) The \xmm\ 2--5 keV image compared with 1.3 GHz image by MeerKAT \citep{Grollimund_2026_v4641_meerkat} and $^{12}$CO($J=1$--0) map by the Nobeyama radio 45-m telescope \citep{hess_v4641_2026}.
The MeerKAT image is shown in units of $\mu$Jy~beam$^{-1}$, and the Nobeyama map is in units of K~\kms.
    \label{fig:MWL}}
\end{figure*}

\addtocounter{figure}{-1}

\begin{figure*}[tb!]
  \centering
  \begin{minipage}{\linewidth}
    \centering
    \includegraphics[width=.8\linewidth]{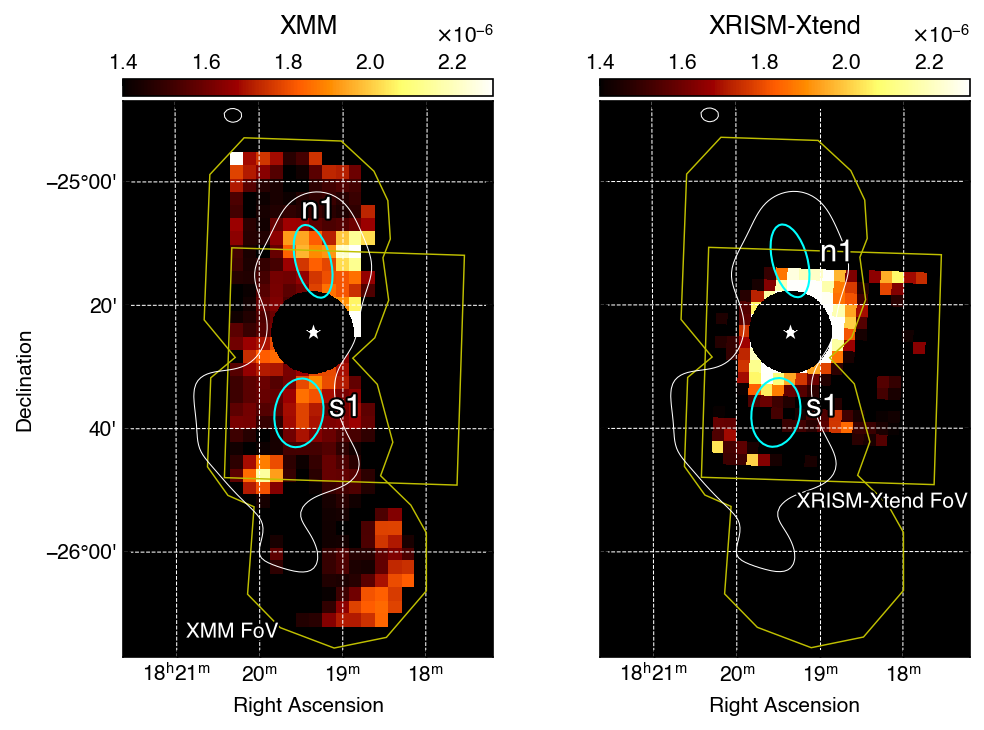}
    \subcaption*{(c) \xmm\ vs \xrism.}
  \end{minipage}
  \caption{
(continued).
(c) Comparison between \xmm\ (left) and \xtend\ \citep[right; ][]{suzukiDetectionExtendedXRay2025}.
Both images are generated in the 1--7 keV band, background-subtracted, and vignetting-corrected, presenting the surface brightness in units of counts~s$^{-1}$. The \xmm\ image is produced without masking point-like sources and is grouped with a coarse binning to match \xtend. The central region ($r = 6.7'$ circle) around \src\ is masked to compare the diffuse emission properly. The bright pixel in the \xmm\ image include bright point sources, which are masked in the other figures.
The white contour shows the $3 \sigma$ significant level in 0.8--22 TeV with \hess\ in (b) and (c).
    }
\end{figure*}

\begin{table}[tb!]
    \begin{center}
    \caption{
    Spatial and spectral information of X-ray knots.
    \label{tab:fit} }
    \begin{tabular}{c | c | c  }
    \hline \hline
         & n1 & s1  \\
    \hline
    R.A. (deg)     & $274.834 \pm 0.001$     & $274.878 \pm 0.002$    \\
    Decl. (deg)    & $-25.205 \pm 0.002$  & $-25.617 \pm 0.002$ \\
    $\sigma_{\rm major}$ (arcmin) & 3.4 $\pm$ 0.1 & 3.3 $\pm$ 0.1 \\
    $\sigma_{\rm minor}$ (arcmin) & 2.1 $\pm$ 0.1 & 2.8 $\pm$ 0.1 \\
    Angle (deg)   &  21.5 $\pm$ 0.1 & $-17.1 \pm 0.1$  \\
    Distance from \src\ (arcmin)  &  12 & 14    \\
    Distance from \src\ (pc)$^{\rm (a)}$ &  23 & 25 \\
    \hline
    \NH\ ($10^{21}$~\column) & $1.7 \pm 0.4_{\rm stat} \pm 0.5_{\rm syst}$    & 1.7 (fixed)   \\
    $\Gamma$                 & $1.9 \pm 0.1_{\rm stat} \pm 0.4_{\rm syst} $ &  $1.8 \pm 0.2_{\rm stat} \pm 0.5_{\rm syst}$  \\
    Flux ($10^{-13}$~\eflux)$^{\rm (b)}$ &  $3.5 \pm 0.5_{\rm stat}  \pm 1.1_{\rm syst}$ & $1.9 \pm 0.5_{\rm stat} \pm 0.6_{\rm syst}$  \\
    Surface brightness ($10^{-15}$~\eflux~arcmin$^{-2}$)$^{\rm (b)}$ &  $6.6 \pm 0.9_{\rm stat}  \pm 2.0_{\rm syst}$  & $6.3 \pm 1.7_{\rm stat}  \pm 1.9_{\rm syst}$   \\
    $\chi^2$ (dof) & 271.2 (243) & 109.3 (72) \\
    \hline
    \end{tabular}
    \end{center}
\tablecomments{
Coordinates, major and minor axes, and position angle (counterclockwise from the north) are obtained by the 2--5 keV image shown in Figure~\ref{fig:map}.
$^{\rm (a)}$ $d=6.2$~kpc is assumed.
$^{\rm (b)}$ Calculated in the 2--10 keV band.
}
\end{table}

\subsection{X-ray Spectra}
\label{sec:spectra}

We extracted X-ray spectra from the cleaned event files (Section~\ref{sec:obs}) after masking point sources (Section~\ref{sec:image}). Redistribution matrix files (RMFs) and auxiliary response files (ARFs) were generated with \texttt{rmfgen} and \texttt{arfgen}, respectively.
To properly account for the spatially extended nature of the emission, we adopted \texttt{extendedsource=yes} and \texttt{detmaptype=flat}, which assumes a uniform detector map when computing the response.
The resulting spectra were grouped with \texttt{grppha} from \texttt{HEASoft} v6.35, requiring a minimum of 100 counts per spectral bin.
We used \texttt{XSPEC} v12.14.1 \citep{1996ASPC..101...17A} for the following spectral fitting.
All errors reported in the text and figures correspond to $1\sigma$ c.l.; unless otherwise mentioned.

We adopt a background-subtraction method.
The source regions are defined in Figure~\ref{fig:map} (right panel), and the background regions are extracted from nearby source-free regions in the same observation:
The background regions are circles centered at (R.A., Decl.) = (18\fh20\fm0\fs3, $-$25\fd12\fm15\fs1) for n1 and (18\fh19\fm23\fs7, $-$25\fd46\fm34\fs0) for s1 with 3.3\arcmin\ and 4.1\arcmin\ in radii, respectively.
We fit the spectra in 0.5--10 keV, ignoring the energy ranges corresponding to the most prominent instrumental fluorescent lines: the 1.2--1.9 keV range (Al and Si lines) for the MOS detectors, and the 1.2--1.7 keV (Al) and 7.0--9.2 keV (Cu) ranges for the PN camera \citep{2024A&A...686A..68R}.
%
The obtained MOS and PN spectra were fitted jointly.

\begin{figure}[tb!]
    \centering
    \includegraphics[width=0.49\linewidth]{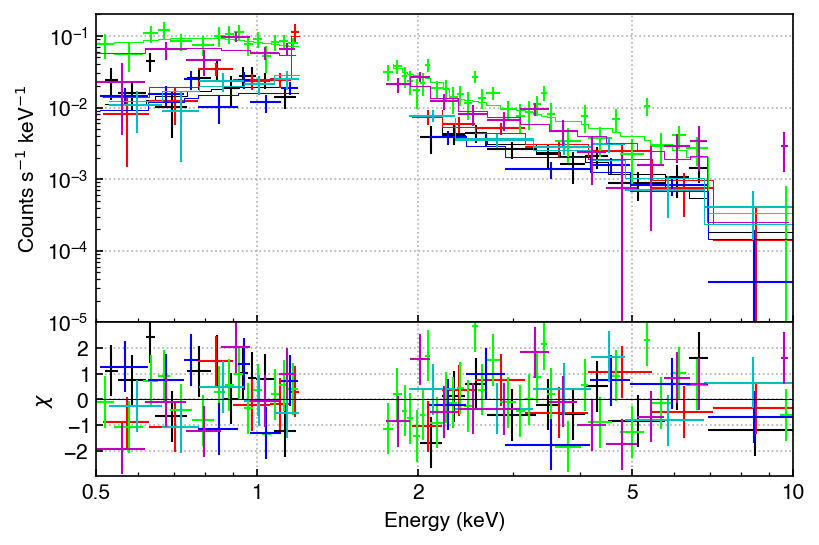}
    \includegraphics[width=0.49\linewidth]{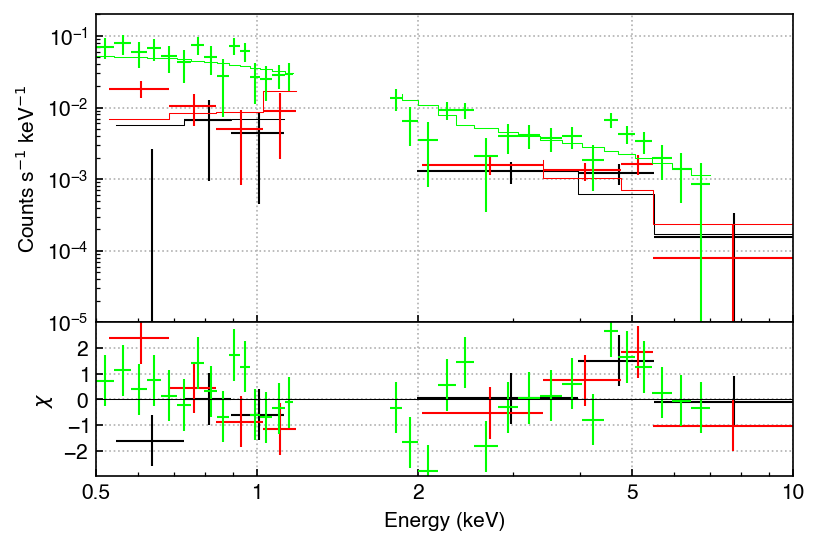}
    \caption{
    X-ray spectra of n1 (left) and s1 (right).
    The MOS1 (black and blue), MOS2 (red and cyan), and PN (lime and magenta) data are simultaneously fitted.
    The solid lines represent the best-fit absorbed power-law model.
    The energy bands contaminated by instrumental line emissions (1.2--1.9 keV for MOS and 1.2--1.7 keV and 7.0--9.2 keV for PN) are ignored (see the text).
    For n1, spectra from two observations 0971600101 and 0971600201 are jointly fitted.
    \label{fig: spectra}}
\end{figure}

The spectra of the n1 and s1 knots are shown in Figure~\ref{fig: spectra}.
They are fitted with an absorbed power-law model (\texttt{tbabs*powerlaw}), adopting the \texttt{wilm} abundance \citep{2000ApJ...542..914W} table for the absorbing column density and cross sections from \cite{1996ApJ...465..487V}.
The best-fit values are summarized in Table~\ref{tab:fit}.
In n1, we obtained a column density of $N_\mathrm{H} = (1.7 \pm 0.4) \times 10^{21}\,\mathrm{cm^{-2}}$, photon index of $\Gamma_X = 1.9 \pm 0.1$, and a 2--10 keV unabsorbed flux of $(4.0 \pm 0.5) \times 10^{-13}$~\eflux.
For s1, the photon index cannot be constrained with $N_\mathrm{H}$ being free; therefore, we fixed the column density to $N_\mathrm{H} = 1.7 \times 10^{21}\,\mathrm{cm^{-2}}$, yielding $\Gamma_X = 1.8 \pm 0.2$ and the 2--10 keV flux of $(1.9 \pm 0.5) \times 10^{-13}$~\eflux.
The detection significance was obtained as $20\sigma$ and $7\sigma$ for n1 and s1, respectively, inferred from the fit statistics with and without the source model.
The obtained surface brightness for n1 and s1 is slightly smaller than that of \xtend\ \citep[][]{suzukiDetectionExtendedXRay2025}, likely due to the contamination of unresolved point sources in \xtend\ (Figure~\ref{fig:MWL}c).
The photon index derived with \xmm\ is roughly in agreement with Case (b) ($\Gamma_X \approx 1.8$) with \xtend, in which a lower background level is assumed in the analysis of the source region than Case (a), resulting in $\Gamma_X \approx 2.4$.
The spectra derived by \xmm\ are roughly consistent with those with EP-FXT \citep{Chi_2026_EP} and about one order of magnitude lower than the upper limits with eROSITA \citep{Bykov_2026_v4641}.
The spectra of the western and southern edge components are poorly constrained; thus, we fixed  $N_\mathrm{H} = 1.7 \times 10^{21}\,\mathrm{cm^{-2}}$ and $\Gamma_X =2$,
and derived surface brightness of
$< 2.0 \times 10^{-15}$~\eflux~arcmin$^{-2}$ and
$(4.3 \pm 2.6) \times 10^{-15}$~\eflux~arcmin$^{-2}$, respectively.
The detection significance was $\sim 1\sigma$ and $\sim 5 \sigma$ for the western and southern edge emission, respectively.

We also tested a possible thermal origin by fitting the spectra with an absorbed collisional ionization equilibrium plasma model (\texttt{tbabs*apec}).
We obtained an electron temperature of $kT \sim 5$--7 keV for n1, with \NH\ being fixed to $1.7 \times 10^{21}$~\column\ and assuming Solar abundance; otherwise, it cannot be constrained.
Compared with the thermal model, the power-law model is significantly preferred ($\Delta$AIC = 88).
Furthermore, the relatively high electron temperature with $kT \gtrsim$~5~keV and no apparent emission lines disfavor a thermal origin.
The emission measure of the \texttt{apec} model was constrained to be
$\int n_e n_{\rm H} dV = (1.1 \pm 0.1)\times10^{56}$~cm$^{-3}$,
corresponding to a plasma density of $\sim$0.07~\cc, assuming a spherical emitting region with a radius of 3\arcmin\ (5.4~pc for $d=6.2$~kpc). This density should be regarded as an upper limit, given that the X-ray emission is likely dominated by a non-thermal component.

The X-ray flux, even an upper limit, places constraints on the magnetic-field strength in the framework of the leptonic scenario.
Since the n1 and s1 knots appear to be more compact than the extended gamma-ray lobes, we derived upper limits on the X-ray flux from the larger gamma-ray-emitting regions after excluding the n1 and s1 regions.
Assuming $\Gamma_X = 2$ and \NH\ $= 1.7 \times 10^{21}$~\column, we obtained 2--10~keV surface-brightness upper limits of $(5$--$6)\times10^{-16}$~\eflux~arcmin$^{-2}$ in the background regions and $(1$--$2)\times10^{-15}$~\eflux~arcmin$^{-2}$ within the 68\% containment regions of the northern and southern gamma-ray components \citep[B1 and B2 in][]{hess_v4641_2026}.

We estimated a systematic uncertainty in the spectral analysis as follows.
We examined that the choice of background region and \NH\ does not largely affect the results.
Although the column density is a key parameter, it remains poorly constrained by the current dataset, except for s1. Reported values in the literature range from $1.6 \times 10^{21}$~\column\ to $6.1 \times 10^{21}$~\column, inferred from the HI data \citep[e.g.,][Web \NH\ calculator\footnote{https://heasarc.gsfc.nasa.gov/cgi-bin/Tools/w3nh/w3nh.pl}]{tomsick_2003_v4641,Doroshenko_2024_3DnHtool}.
We also tested the fitting energy range in the 1--10~keV band, excluding the soft band which is contaminated by a spatially non-uniform sky component. 
%
We concluded that the spectral parameters are consistent within $\sim$30\% of the best-fit values listed in Table~\ref{tab:fit}, depending on the background regions, adopted \NH\ value, and fitting energy range. 

\subsection{SED modeling} \label{sec:modeling}

Here we present broadband spectral energy distribution (SED) modeling from X-rays to gamma rays.
Since the X-ray knots ($\sigma_{\rm major} \approx 3\arcmin$ and $\sigma_{\rm minor} \approx 2 \arcmin$) are smaller than the TeV lobes \citep[$\sigma_{\rm major} \approx 9$--$11\arcmin$ and $\sigma_{\rm minor} \approx 2$--$7\arcmin$ in the two-component model;][]{hess_v4641_2026},
we introduced a volume-filling factor ($f$) representing the fraction of the gamma-ray-emitting volume occupied by the X-ray knots.
Although the difference in sizes seems significant even accounting for the different PSFs of the instruments, we note that this should not be overinterpreted as the gamma-ray extent is more uncertain in the two-component model.
The filling factor is 0.045--0.16 for n1 and 0.025--0.04 for s1, estimated from the projected angular size and an assumed line-of-sight depth. These ranges were obtained by adopting the line-of-sight depth to be (1) $\sigma_{\rm major}$, (2) $\sigma_{\rm minor}$, and (3) the geometric mean of the major- and minor-axis diameters of the 68\% containment region.
When we fit the X-ray spectra of n1 and s1, the gamma-ray SED of the northern and southern lobes are scaled by the filling factor, respectively.

\begin{figure}[tb!]
    \centering
    \includegraphics[width=\linewidth]{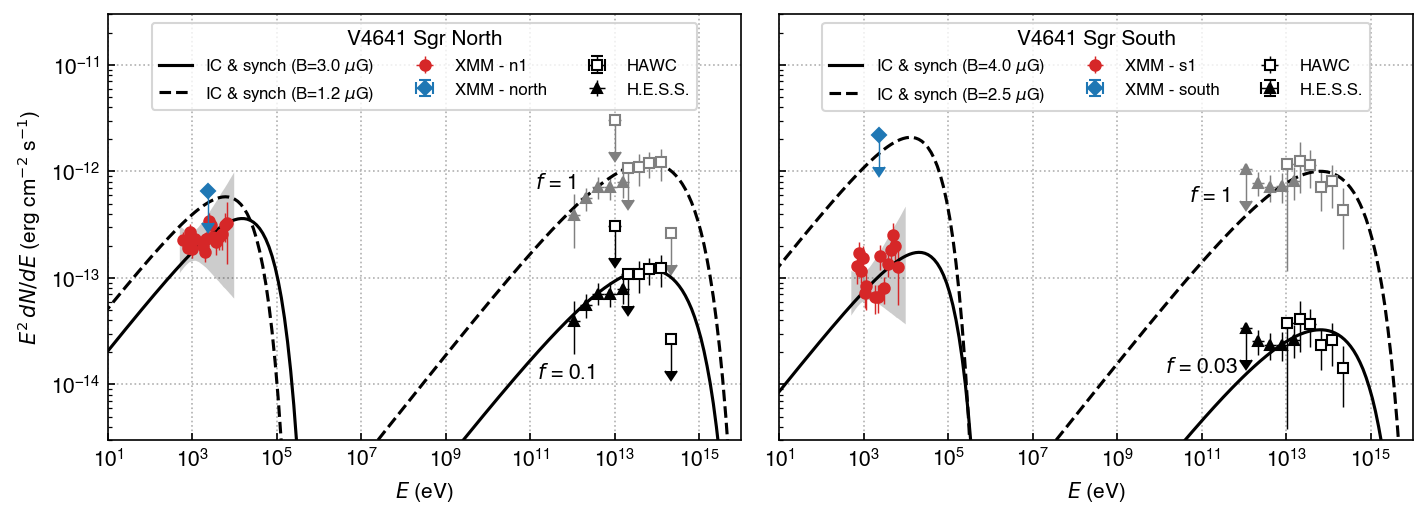}
    \includegraphics[width=\linewidth]{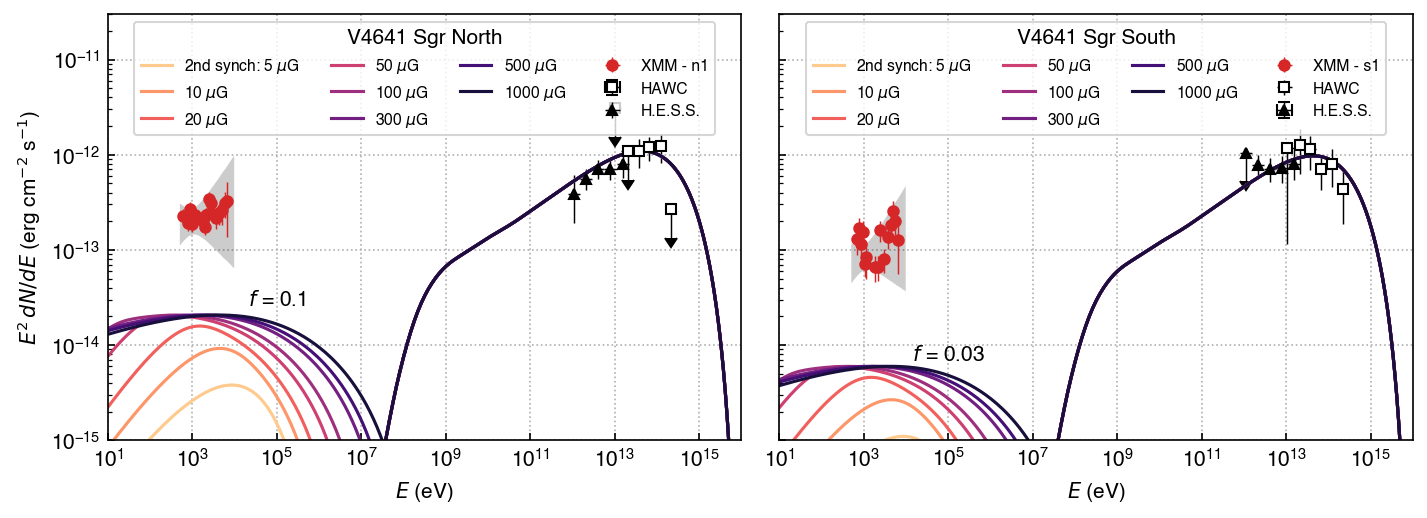}
    \caption{
    Broadband SED of the north lobe (left) and the south lobe (right).
    Top: the leptonic modeling.
    The \xmm\ spectra are corrected for the absorption with \NH\ of $1.7 \times 10^{21}$~\column. The systematic uncertainty is shown with the gray shade.
    The X-ray upper limits (in blue) are scaled to the 95\% containment region by \hess\
    The gamma-ray spectra (in gray) are adopted from measurements by HAWC and \hess\ \citep{hawc_2024_v4641,hess_v4641_2026}.
    The black gamma-ray spectra are scaled by the filling factor ($f$; see the text) to compare with the n1 and s1 spectra.
    The solid and dashed lines indicate the leptonic models for the X-ray knots (n1 and s1) and the non-X-ray regions, respectively.
    Bottom: the hadronic modeling.
    Only the models of the secondary synchrotron components are scaled by $f$, while the other spectra and model are shown without any scaling.
    The hadronic model is calculated for $n = $1~\cc.
    \label{fig:SED}}
\end{figure}

In the leptonic case, we assumed a cutoff power-law electron distribution, $dN_e/dE_e \propto E_e^{-s_e}\, \exp(-E_e/E_{e,{\rm max}})$ with $s_e = 2$ and $E_{e, {\rm max}} = 1$~PeV \citep{hawc_2024_v4641,hess_v4641_2026}.
The gamma-ray emission was fitted with inverse-Compton scattering using the photon fields reported in \cite{hess_v4641_2026}, while the X-ray emission was modeled as synchrotron radiation from the same electron population with the magnetic field strength, $B$, being a free parameter.
The inferred magnetic-field is $B \approx 3$--5~\uG\ for n1 and s1 (Figure~\ref{fig:SED}).
On the other hand, we derived $B \lesssim 1$--2~\uG\ for the non-X-ray-detected part of the northern and southern lobes (i.e., the \hess\ B1 and B2 regions after excluding n1 and s1), as shown in Figure~\ref{fig:SED}. Here the measured X-ray upper limits on the surface brightness (Section~\ref{sec:spectra}) were scaled to the gamma-ray emission size, thus the filling factor was not adopted.
Note that the radio fluxes by MeerKAT \citep{Grollimund_2026_v4641_meerkat} are considered upper limits since there is no apparent counterpart of the X-ray and gamma-ray lobe in the MeerKAT observation (Figure~\ref{fig:MWL}), and they are consistent with our modelings.
%

In the hadronic scenario, synchrotron X-rays can be emitted from either primary electrons or secondary electrons and positrons produced through proton-proton interactions \citep[e.g.,][]{aharonian_gamma_2013,tsuji_search_2024}.
In both cases, the strength of the magnetic field can be significantly larger than the value inferred above.
Although the hadronic-dominant origin seems disfavored by the unrealistically large proton power required and the absence of dense gas \citep[Figure~\ref{fig:MWL}b;][]{hess_v4641_2026}, we calculated the hadronic gamma-ray and secondary synchrotron emission assuming a cutoff power-law proton distribution of $s_p =1.8$ and $E_{p,{\rm max}} = 3$~PeV, as shown in Figure~\ref{fig:SED}.
Protons were injected continuously for 1 kyr at a constant rate. The resulting secondary electrons were evolved self-consistently under radiative cooling in a magnetic field ranging from $B$ = 5 to $10^3$~\uG, which represents the magnetic field in the X-ray-emitting knot (Figure~\ref{fig:SED}).
When the observed volume filling factor of the X-ray knots is taken into account, the predicted secondary synchrotron emission falls below the observed X-ray fluxes (Figure~\ref{fig:SED}).
This reflects the simple assumption that the secondaries uniformly occupy the same proton distribution (i.e., the gamma-ray-emitting region) and the compact X-ray knots trace stronger magnetic fields.
If the filling-factor correction is not applied, the expected secondary synchrotron flux increases by a factor of $1/f$ ($\sim$ 10--30), becoming comparable to the observed X-ray emission.
In reality, more realistic two-zone numerical modeling incorporating spatially varying magnetic fields and particle transport, together with deeper X-ray observations to better determine the intrinsic extent of the knots and search for fainter diffuse emission, will be required to assess the viability of this scenario.

\section{Discussion} \label{sec:discussion}

Together with the Einstein Probe observation \citep{Chi_2026_EP},
we report the discovery of two extended X-ray features, n1 and s1, symmetrically located at the north and south of \src\ and aligned with both the radio-jet axis and the TeV lobes   \citep{hawc_2024_v4641,Marti_2026_v4641,hess_v4641_2026,lhaaso_2025_microquasars}.
These features provide the first evidence for parsec-scale X-ray jet structures in this microquasar, and such features are known only in several systems, such as jets in SS~433, XTE J1550$-$564, H~1743$-$322, 4U~1755$-$33, Cir X-1, and MAXI~J1820+070 \citep{safi-harbROSATASCAObservations1997,safi-harbRossiXRayTiming1999,migliariIronEmissionLines2002,sakaiArcsecondScaleXrayImaging2025,Corbel_2002_XTEJ1550,Corbel_2005_h1743,Angelini_2003_4U1755,Kaaret_2006_4u1755,Heinz_2007_CirX1,Espinasse_2020_MAXIJ1820}.
Remarkably, the broadband SED is dominated by gamma rays rather than X-rays, the X-ray extent appears smaller than the TeV-emitting region, and the X-ray morphology is diffuse unlike the sharp shock-like structures in the innermost knots and subsequent flow-like structures in the downstream regions observed in SS~433/W50.
Here we propose two possible interpretations: (1) n1 and s1 correspond to the innermost acceleration sites, such as termination or (re-)collimation shocks powered by the jet activities, or (2) they are downstream structures.
In either case, the magnetic field strength of n1 and s1 is inferred to be only 3--5~\uG, yet is likely locally amplified compared with the surrounding gamma-ray-emitting region with $B \lesssim 1$--2~\uG\ (Section~\ref{sec:modeling}), supporting a scenario in which localized magnetic-field amplification contributes to the formation of the n1 and s1 regions \citep[e.g.,][]{sudohMultiwavelengthEmissionGalactic2020}.

A useful comparison can be made with the SS~433/W50 system. The projected offsets of n1 and s1 from \src, $\sim20$ pc, are comparable to those of the innermost knots in the outer jets of SS~433/W50 \citep{safi-harbROSATASCAObservations1997,safi-harbHardXRayEmission2022a}, suggesting a similar physical origin.
However, a key difference is the gamma-ray-dominated SED of \src\ with $F_X/F_\gamma \sim 0.1$ ($\sim$ 1--3 for the filling factor $f \approx$ 0.03--0.1), whereas SS~433/W50 exhibits $F_X/F_\gamma \sim 10$.
This difference reflects the substantially weaker magnetic field in \src\ ($B\sim 3$--5~\uG) compared with SS~433/W50 ($B\gtrsim10$--20~\uG$;$ \citealt{safi-harbHardXRayEmission2022a,kayamaSpatiallyResolvedStudy2022,kayamaXrayStudyPropagation2025,hess_2024_ss433}). The weaker field may result from the lower time-averaged jet power of \src\ and/or its different environment, as \src\ lies outside the Galactic plane ($b \approx -5$\degr) and may interact with a more tenuous ISM.
This weak magnetic field naturally leads to larger gyroradii and may contribute to the more diffuse morphology of n1 and s1 than the innermost knots in SS~433/W50.
Alternatively, if the shear acceleration process is responsible for the non-thermal X-ray emission detected \citep[e.g.,][]{Wan_2026_v4641}, any sharp feature is not expected.
Additionally, n1 and s1 have offsets of $\sim16$\degr\ and $\sim10$\degr, respectively, from the extrapolated radio jet axis (Figure~\ref{fig:map}), which is different from almost perfect alignment of X-ray knots in the radio jet in SS~433 \citep{Sunyaev_2026_ss433}.
This offset is comparable to the reported misalignment of $\gtrsim12$\degr\ between the jet axis and the binary orbital axis \citep{Marti_2026_v4641}.
The X-ray knots may therefore trace the time-averaged jet orientation rather than the instantaneous jet direction inferred from the radio observations,
or this misalignment may be due to the unknown precession of the jet.
Furthermore, the major axis of n1 is inclined by $\sim45^\circ$ relative to the jet direction, and its relatively soft spectrum ($\Gamma_X\sim1.9$) is more reminiscent of downstream regions such as e2 than of the innermost knots in SS~433. These properties may indicate that n1 and s1 are not direct termination shocks but instead represent downstream structures.
Future X-ray observations may reveal additional knots outside the current FoVs, such as e3 in SS~433/W50 as known as the radio ear or a thermal knot.


The observed weak magnetic field (Section~\ref{sec:modeling}) has important implications on the X-ray morphology of n1 and s1, which is likely determined by a combination of particle acceleration, transport, magnetic fields, and radiative cooling.
If n1 and s1 represent the acceleration sites, acceleration to the PeV energies requires a jet velocity of at least $\sim$0.1$c$ ($\sim$30,000~km~s$^{-1}$) based on the Hillas criterion \citep{Hillas_1984,hess_v4641_2026}, adopting $B \sim $3--5~\uG\ and the acceleration site size of $R \sim 5$~pc.
With $v_{\rm jet} \gtrsim 0.1 c$, the advection timescale ($t_{\rm adv} \approx R/v_{\rm jet}$) is substantially shorter than the radiative cooling time, which is dominated by the synchrotron process with $B \sim$ 3--5~\uG\ \citep{hess_v4641_2026}.
Thus, radiative losses alone are unlikely to determine the observed knot size.
This conclusion remains unchanged for moderate post-shock deceleration ($v_{\rm down}\sim v_{\rm sh}/4$), as expected for a strong shock.
However, if the downstream flow is further decelerated to $\lesssim 0.01c$, radiative cooling could become effective in determining the knot extent.
Alternatively, the X-ray morphology may also be regulated by spatial and temporal decay of an amplified turbulent magnetic field downstream of multiple shocks \citep{Bykov_2025_ss433,Bykov_2026_v4641}.
If particle transport in the X-ray knots is dominated by diffusion, the characteristic diffusion length can be expressed as
$R_{\rm d} = 2 \sqrt{D(E_e) t_{\rm rad}}$,
where $D(E_e)=D_{100}(E_e/100~{\rm TeV})^\delta$ is the diffusion coefficient with $\delta$ = 1/3--1.
We obtained $D_{100}\approx5\times10^{26}$~cm$^2$~s$^{-1}$ for $B = 4~\uG$, and $R_d = 5$~pc. This value is close to the Bohm limit and approximately three orders of magnitude below the typical ISM diffusion coefficient \citep{Gabici_2019_cr}, indicating strongly suppressed diffusion and highly efficient particle acceleration around the knots.
This interpretation implies, however, a very slow bulk speed of the flow, \(v_{\mathrm{jet}}\ll0.01c\).

The inferred energetics are also reasonable.
The observed X-ray luminosity from n1 and s1 is $\sim 10^{33}$~erg~s$^{-1}$, derived from the X-ray flux of $\sim $(2--4)$\times 10^{-13}$~\eflux\ and $d=6.2$~kpc.
Accounting for the modest efficiency of the synchrotron emission due to the fast advection, \(\lesssim10\%\), and the fraction of jet energy transferred to electrons, \(\sim0.1\%\), inferred in SS~433, we should expect that the jet power in \src\, should be relatively high, $\gtrsim 10^{37}$~erg~s$^{-1}$.
Although the 1999 outburst reached a super-Eddington luminosity exceeding $10^{39}$~erg~s$^{-1}$, the luminosity of the central black hole can occasionally drop to as low as $10^{31}$~erg~s$^{-1}$; with the averaged luminosity being estimated to be in the range of $10^{35}$--$10^{37}$~erg~s$^{-1}$ \citep{tomsick_2003_v4641,2026arXiv260316714L}.
Thus, the X-ray luminosity favors the upper end, although still in the range anticipated, for the jet averaged luminosity. Furthermore, the TeV gamma-ray luminosity, $\sim5\times10^{34}$~erg~s$^{-1}$, serve as a better proxy for the energy transferred to relativistic electrons, which is consistent with a few $\times10^{-3}$ of the jet power being transferred to relativistic electrons for a jet power of $\sim 10^{37}$~erg~s$^{-1}$.

A hadronic origin of the TeV emission remains viable.
Although the observed X-ray is unlikely originated from synchrotron emission of the secondaries with the measured filling factor (Section~\ref{sec:modeling} and Figure~\ref{fig:SED}), the hadronic case would reproduce the diffuse morphologies of n1 and s1, since the X-ray emission traces a combination of the proton interaction rate and the magnetic field strength.
As mentioned in Section~\ref{sec:modeling}, more detailed calculations are required, which is beyond the scope of this paper.
An alternative possibility is a hybrid scenario, in which the gamma rays are produced by protons, while the observed X-rays originate from synchrotron radiation of primary electrons accelerated in localized regions of amplified magnetic fields. Such a picture allows a much stronger magnetic field strength than in pure leptonic case and naturally explains the compact X-ray knots embedded within more extended gamma-ray lobes, owing to the much shorter cooling timescale of TeV electrons compared with PeV protons.
Searching for a hadronic component remains important in the context of the origin of Galactic cosmic rays.
Even in a predominantly leptonic PeVatron system such as \src, protons may still be accelerated up to energies of at least $\sim$1~PeV. Such hadronic signatures could be probed by future observations of gas, secondary synchrotron X-rays, and gamma rays with higher angular resolution using facilities such as ALMA, HEROIX\footnote{\url{https://sti.usra.edu/missions/heroix/}}, and CTAO, respectively.


The scheduled follow-up observations with \chandra\ and \nustar\ will further constrain the morphology and spectra of the diffuse X-ray features.
Although the X-ray emission spatially correlates with the TeV lobes, we cannot completely exclude the possibility that it originates from unresolved point sources.
Also, we reported the apparent diffuse morphology, which, however, may result from projection effects, smoothing effects, or the limited angular resolution of the current observations.
The superior angular resolution of \chandra\ ($\sim$0.5$^{\prime\prime}$) will be crucial for testing the truly intrinsic diffuse nature of n1 and s1 (as well as the marginally detected western and further southern components) and identifying possible point-source contamination.
\nustar\ observations will probe the hard X-ray band up to 79~keV, confirming the thermal/non-thermal nature of the X-ray regions and potentially constraining the maximum leptonic particle energy in the system.
Detailed modeling, including numerical simulation, of particle transport and radiation
will further unveil the nature of n1 and s1.


In this Letter, we reported the discovery of localized extended X-ray emission coincident with the gamma-ray emission in both the northern and southern lobes of \src, aligned with the radio-jet axis (although with a slight offset).
These largely extended jets at parsec-scale have been identified in only a few X-ray binaries.
In particular, the similarity between the X-ray knots in \src\ and the SS~433/W50 system suggests that they play crucial roles in particle acceleration in microquasar environments.
However, such structures have likely been overlooked in previous studies, which primarily focused on the bright central source during outburst phases. Emission from the central binary can dominate the FoVs of the X-ray satellites and obscure faint extended features in the surrounding regions.
This \xmm\ study demonstrated the importance of deep X-ray observations of spatially offset regions during quiescent phases. Similar observations of other microquasars may reveal additional X-ray knots, potentially tracing particle-acceleration sites and/or enhanced magnetic fields associated with gamma-ray radiation and relativistic jets.

\begin{acknowledgments}

We thank the anonymous referee for helpful comments.
We also thank
Tong Bao,
Ruo-yu Liu,
Alberto Moretti,
Giovanni Pareschi, and
Hiroki Yoneda for the fruitful discussion and preparation for the X-ray proposals.
NT acknowledges support by the Japan Society for the Promotion of Science (JSPS) KAKENHI grant Nos. JP22K14064 and JP26H00820. KM and ML are supported by NASA's NuSTAR AO-11 grant (80NSSC26K0044), which was awarded for the XMM-Newton and NuSTAR survey program of V4641 Sgr.
DK acknowledges support by RSF grant No. 24-12-0045.
SM acknowledges support from grants NASA ADAP 80NSSC24K0666 and NASA NuSTAR data analysis funding 1729326.
EdelaF acknowledges support from the ICRR Inter-University Research Programme (grant 2026i-F-007), carried out within the ICRR international joint research program.

\end{acknowledgments}

\begin{contribution}

NT was responsible for the data analysis and writing and submitting the manuscript.
CT was responsible for the data analysis and writing the manuscript.
KM managed the entire X-ray observation project of V4641 Sgr and edited and reviewed the manuscript.
DK was responsible for the interpretation.
HS provided the formal analysis and validation of comparison with the XRISM-Xtend data.
LON was provided the validation of comparison with the gamma-ray data.
SSH and ML were responsible for writing and reviewing the manuscript.
GP provided CT with the tools, guidance, and supervision to perform the \xmm\ analysis.
All other authors contributed to drafting the observation proposals, the discussion, and reviewing the manuscript.


\end{contribution}

%
\facilities{
\xmm\ (EPIC)
}

\software{
HEASoft v6.35,
SAS v22.1.0,
XSPEC v12.14.1 \citep{1996ASPC..101...17A}
          }

\appendix



\section{Summary of X-ray Observations}
\label{appendix: X-ray data set}

Table~\ref{table:obs} summarizes the \xmm\ observations analyzed in this work.

\begin{table*}[ht!]
\caption{Observation log}
\label{table:obs} 
\centering
\begin{tabular}{c c c c c c}      
\hline\hline               
ObsID & Date & Instrument & Effective Exposure & R.A. & Decl.\\
    & & & (ks) & (deg) & (deg) \\         
\hline                      
    0971600101 & 2025-09-11 & MOS1 & 38 & 274.87 & $-$25.113 \\
     &  & MOS2 &  38\\
     &  & PN &  33\\
     \hline
    0971600201 & 2025-09-13 & MOS1 & 42 & 274.88 & $-$25.280 \\
     &  & MOS2 &  42\\
     &  & PN &40\\
     \hline
    0971600301 & 2025-09-21 & MOS1 & 33 & 274.87 & $-$25.671 \\
     &  & MOS2 &  33\\
     &  & PN &  29\\
     \hline
    0971600401 & 2025-09-23 & MOS1 & 14 & 274.75 & $-$26.017 \\
     &  & MOS2 & 14\\
     &  & PN  & 8\\
\hline                  
\end{tabular}
\end{table*}

\bibliography{reference,reference2}
\bibliographystyle{aasjournalv7}



\end{CJK*}
\end{document}